\documentclass[acmsmall,screen]{acmart}

\usepackage{ragged2e}
\usepackage{multirow}
\usepackage{diagbox}
\usepackage{color}
\usepackage{censor}
\usepackage{url}
\usepackage{booktabs}
\usepackage{pifont}
\usepackage{setspace}
\newtheorem{definition}{Definition}
\newcommand{\toolname}{RetriGen}
\usepackage{xspace}
\usepackage{url}
 
\newcommand{\ie}{\textit{i.e.,}\xspace}
\newcommand{\eg}{\textit{e.g.,}\xspace}

\newcommand{\etal}{\textit{et al.}\xspace}

\newcommand{\newdata}{$Data_{new}$\xspace}
\newcommand{\olddata}{$Data_{old}$\xspace}

\newcommand{\irar}{$IR_{ar}$\xspace}
\newcommand{\rah}{$RA_{adapt}^{H}$\xspace}
\newcommand{\rann}{$RA_{adapt}^{NN}$\xspace}
\newcommand{\inte}{\textit{Integration}\xspace}
\newcommand{\atla}{\textit{ATLAS}\xspace}
\newcommand{\edit}{\textsc{EditAS}\xspace}

\colorlet{BLUE}{blue} 
\colorlet{ORANGE}{orange}
\colorlet{BLACK}{black}
\usepackage[normalem]{ulem}
\newcommand{\revise}[1]{{\color{black}{#1}}}
\newcommand{\delete}[1]{}

\usepackage{subfigure}
\usepackage{colortbl}
\usepackage{balance}

\usepackage[most]{tcolorbox}
\newcommand{\finding}[2]{
\begin{center}
\begin{tcolorbox}[colback=gray!15, colframe=black, boxsep= -0.15cm, middle=-0.15cm, breakable]
\textbf{Answer to RQ{#1}:}
{#2}
\end{tcolorbox}
\end{center}
}

\usepackage[most]{tcolorbox}
\newcommand{\find}[1]{
\begin{tcolorbox}[leftrule=1mm,toprule=0mm,bottomrule=0mm,left=1pt,right=2pt,top=2pt,bottom=2pt,breakable]%
\em #1
\end{tcolorbox}
}

\AtBeginDocument{
  \providecommand\BibTeX{{
    \normalfont B\kern-0.5em{\scshape i\kern-0.25em b}\kern-0.8em\TeX}}}

\setcopyright{acmcopyright}
\copyrightyear{2025}
\acmYear{2025}
\acmDOI{1}

\acmJournal{TOSEM}
\acmVolume{1}
\acmNumber{1}
\acmArticle{1}
\acmMonth{1}

\begin{document}

\title{Improving Deep Assertion Generation via Fine-Tuning Retrieval-Augmented Pre-trained Language Models}

\author{Quanjun Zhang} 
\orcid{0000-0002-2495-3805}
\email{quanjun.zhang@smail.nju.edu.cn}

\affiliation{
  \institution{Department of Computing Technologies, Swinburne University of Technology}
  \city{Melbourne}
  \country{Australia}
}

\affiliation{
  \institution{State Key Laboratory for Novel Software Technology, Nanjing University}
  \city{Nanjing}
  \state{Jiangsu}
  \country{China}
}

\author{Chunrong Fang} 
\authornote{\textbf{Chunrong Fang, Rubing Huang and Zhenyu Chen are the corresponding authors.}}
\orcid{0000-0002-9930-7111}
\email{fangchunrong@nju.edu.cn}

\author{Yi Zheng} 
\orcid{0000-0002-2495-3805}
\email{201250182@smail.nju.edu.cn}

\author{Yaxin Zhang} 
\orcid{0000-0002-2495-3805}
\email{zhangyaxin032@gmail.com}

\author{Yuan Zhao} 
\orcid{0000-0002-2495-3805}
\email{allenzcrazy@gmail.com}

\affiliation{
  \institution{State Key Laboratory for Novel Software Technology, Nanjing University}
  \city{Nanjing}
  \state{Jiangsu}
  \country{China}
}

\author{Rubing Huang} 
\authornotemark[1]
\orcid{0000-0002-2495-3805}
\email{rbhuang@must.edu.mo}

\affiliation{
  \institution{School of Computer Science and Engineering, Macau University of Science and Technology}
  \city{Nanjing}
  \state{Jiangsu}
  \country{China}
}

\author{Jianyi Zhou}
\email{zhoujianyi2@huawei.com}
\orcid{}
\affiliation{
  \institution{Huawei Cloud Computing Technologies Co., Ltd.}
  \city{Beijing}
  \country{China}
}

\author{Yun Yang}
\email{yyang@swin.edu.au}
\orcid{0000-0002-7868-5471}
\affiliation{
  \institution{Department of Computing Technologies, Swinburne University of Technology}
  \city{Melbourne}
  \country{Australia}
}

\author{Tao Zheng}
\email{zt@nju.edu.cn}
\orcid{0000-0002-9592-7022}

\author{Zhenyu Chen}
\authornotemark[1]
\email{zychen@nju.edu.cn}
\orcid{0000-0002-9592-7022}

\affiliation{
  \institution{State Key Laboratory for Novel Software Technology, Nanjing University}
  \city{Nanjing}
  \state{Jiangsu}
  \country{China}
}
\affiliation{
  \institution{Shenzhen Research Institute of Nanjing University}
  \city{Shenzhen}
  \state{Guangdong}
  \country{China}
}

\renewcommand{\shortauthors}{Trovato and Tobin, et al.}

\begin{abstract}
Unit testing validates the correctness of the units of the software system under test and serves as the cornerstone in improving software quality and reliability.
To reduce manual efforts in writing unit tests, some techniques have been proposed to generate test assertions automatically, including deep learning (DL)-based, retrieval-based, and integration-based ones.
Among them, recent integration-based approaches inherit from both DL-based and retrieval-based approaches and are considered state-of-the-art.
Despite being promising, such integration-based approaches suffer from inherent limitations, such as retrieving assertions with lexical matching while ignoring meaningful code semantics, and generating assertions with a limited training corpus.

In this paper, we propose a novel \textbf{Retri}eval-Augmented Deep Assertion \textbf{Gen}eration approach, namely {\toolname}, based on a hybrid assertion retriever and a pre-trained language model (PLM)-based assertion generator.
Given a focal-test, \toolname{} first builds a hybrid assertion retriever to search for the most relevant test-assert pair from external codebases.
The retrieval process takes both lexical similarity and semantical similarity into account via a token-based and an embedding-based retriever, respectively.
\toolname{} then treats assertion generation as a sequence-to-sequence task and designs a PLM-based assertion generator to predict a correct assertion with historical test-assert pairs and the retrieved external assertion.
Although our concept is general and can be adapted to various off-the-shelf encoder-decoder PLMs, we implement \toolname{} to facilitate assertion generation based on the recent CodeT5 model.
We conduct extensive experiments to evaluate \toolname{} against six state-of-the-art approaches across two large-scale datasets and two metrics.
The experimental results demonstrate that {\toolname} achieves 57.66\% and 73.24\% in terms of accuracy and CodeBLEU, outperforming all baselines with an average improvement of 50.66\% and 14.14\%, respectively.
Furthermore, \toolname{} generates 1598 and 1818 unique correct assertions that all baselines fail to produce, 3.71X and 4.58X more than the most recent approach \edit{}.
We also demonstrate that adopting other PLMs can provide substantial advancement, \eg four additionally-utilized PLMs outperform \edit{} by 7.91\%$\sim$12.70\% accuracy improvement, indicating the generalizability of \toolname{}. 
Overall, our study highlights the promising future of fine-tuning off-the-shelf PLMs to generate accurate assertions by incorporating external knowledge sources.

\end{abstract}



\maketitle

\section{Introduction}
\label{sec:intro}
Unit testing has become a pivotal and standard practice in software development and maintenance, serving as a fundamental phase in improving software quality and reliability~\cite{dinella2022toga, liu2023towards}.
This practice involves writing unit tests to ensure that individual components (\eg methods, classes, and modules) perform as per their designated specifications and usage requirements.
Unlike integration and system testing~\cite{arcuri2019restful}, which evaluate the system's overall functionality, unit testing is dedicated to ensuring that individual components of the system operate as intended by the developer, enabling early detection and diagnosis of issues during software development~\cite{schafer2023empirical}.
Besides, well-designed unit tests enhance the quality of production code, minimize the costs of software failures, and facilitate debugging and maintenance processes~\cite{hartman2002issta,elbaum2002test,planning2002economic}.

Despite the significant benefits of unit testing, it is non-trivial and time-consuming to construct effective unit tests manually.
For example, as shown in a prior report~\cite{daka2014survey}, software developers usually spend more than 15\% of their time on writing unit test cases.
Thus, a substantial amount of research has been dedicated to automated unit test generation, such as Randoop~\cite{pacheco2007randoop} and EvoSuite~\cite{fraser2011evosuite}.
A typical unit test usually consists of two key components: (1) the test prefix involving a sequence of statements that configure the unit under test to reach a particular state, and (2) the test oracle containing an assertion that defines the expected outcome in that state~\cite{dinella2022toga}.
Nevertheless, these test generation tools often focus on creating tests that achieve high coverage, while struggling to understand the intended program behavior and produce meaningful assertions.
For instance, Almasi~\etal~\cite{almasi2017industrial} conduct an industrial evaluation of unit tests generated by EvoSuite, highlighting that assertions in manually written tests, in contrast to automatically generated ones, are more meaningful and practical.

To tackle the assertion problem suffered by existing unit test generation tools, an increasing number of assertion generation (AG) techniques~\cite{watson2020learning,yu2022automated,sun2023revisiting} have been proposed.
Such AG techniques can be categorized into three groups: deep learning (DL)-based, retrieval-based, and integration-based ones.
In particular, DL-based AG techniques (\eg \atla{}~\cite{watson2020learning}) handle the assertion generation problem as a neural machine translation (NMT) task with sequence-to-sequence learning.
For example, \atla{}~\cite{watson2020learning} takes a focal method (\ie the method under test) and its test prefix as input to generate an assertion from scratch.
For consistency with prior research~\cite{sun2023revisiting}, we refer to the input as \emph{focal-test}, and the input-output pair as a \emph{test-assert pair}.
However, DL-based AG techniques are restricted by the quality and amount of historical test-assert pairs for training, \eg a small corpus with only 156,760 samples for \atla{}.
In contrast to DL-based AG techniques, given an input focal-test, retrieval-based AG techniques (\eg \irar{}, \rann{} and \rah{}~\cite{yu2022automated}) retrieve the most similar focal-test from a codebase and adapt its assertion to obtain the desired outcome.
However, retrieval-based AG techniques face difficulties in retrieving accurate assertions based on only lexical similarity, which is sensitive to the choice of the identifier naming of source code while ignoring the meaningful code semantics.

Recently, integration-based AG techniques (\eg \inte{}~\cite{yu2022automated} and \edit{}~\cite{sun2023revisiting}) leverage both DL-based and retrieval-based approaches as basic components to retrieve or generate assertions.
For example, \inte{} first utilizes \irar{} to retrieve similar assertions, and use \atla{} to generate new assertions from scratch for those deemed appropriate.
Meanwhile, \edit{} trains a neural model to generate edit actions for similar assertions retrieved by \irar{}.
Despite their impressive performance, integration-based AG techniques still inherit the limitations of its two components, \ie retrieving assertions by lexical similarity while overlooking code semantics, and generating assertions with a small training corpus.

In this paper, we propose a novel retrieval-augmented AG approach called {\toolname} equipped with a hybrid assertion retriever and a pre-trained language model (PLM)-based assertion generator to address limitations of prior work.
Our work is inspired by the opportunity to integrate the well-known plastic surgery hypothesis~\cite{barr2014plastic} with recent PLMs~\cite{wang2021codet5} in the field of assertion generation.
To this end, we automate the plastic surgery hypothesis by fine-tuning retrieval-augmented PLMs, \ie retrieving similar assertions from open-source projects to assist in fine-tuning PLMs for new assertion generation.
Particularly, given a focal-test, {\toolname} builds a hybrid retriever to jointly search for the most relevant focal-test and its assertion from external codebases.
The hybrid retriever consists of a sparse token-based retriever and a dense embedding-based retriever to consider both the lexical and semantic similarity of test-assert pairs in a unified manner.
{\toolname} then uses off-the-shelf PLMs as a backbone of the generator to perform the assertion generation task by fine-tuning it with the focal-test and the retrieved external assertion as input.
{\toolname} is conceptually generalizable to various encoder-decoder PLMs, and we implement {\toolname} on top of the recent code-aware CodeT5 model.
While the retrieval-augmented generation pipeline has been explored in previous code-related studies~\cite{wang2023rap, nashid2023retrieval, parvez2021retrieval}, we are the first to investigate its effectiveness for assertion generation by utilizing a hybrid retriever to fine-tune off-the-shelf PLMs with external knowledge sources.
The distinctions between \toolname{} and previous AG approaches mainly lie in both the retriever and the generator.
First, prior work utilizes an assertion retriever (\eg Jaccard similarity~\cite{sun2023revisiting}) based on lexical matching, while \toolname{} builds a hybrid retriever to search for relevant assertions jointly, demonstrating superiority over a single retriever.
Besides, prior work trains an assertion generator with \delete{a vanilla transformer}\revise{a basic encoder-decoder model} (\eg RNNs~\cite{watson2020learning}) from a limited number of labeled training data, 
while \toolname{} is built upon off-the-shelf language models, which are pre-trained from various open-source projects in the wild to obtain general knowledge about programming language, thus generating optimal vector representations for unit tests.

We select six state-of-the-art AG approaches as baselines, including one DL-based approach (\ie \atla{}), three retrieval-based approaches (\ie \irar{}, \rah{}, and \rann{}), two integration-based approaches (\ie \inte{} and its most recent follow-up \edit{}).
We conduct extensive experiments to compare {\toolname} with these baselines on two widely adopted datasets \newdata{} and \olddata{} in terms of both accuracy and CodeBLEU.
The experimental results show that \toolname{} outperforms all baselines across all datasets and metrics, with average accuracy and CodeBLEU of 57.66\% and 73.24\%, setting new records in the AG field.
The average improvement against these baselines in accuracy and CodeBLEU score reached 50.66\% and 14.14\% on the two datasets.
Besides, \toolname{} generates 1598 and 1818 unique assertions that all baselines fail to predict, significantly improving the most recent approach \edit{} by 3.71X and 4.58X, respectively,  demonstrating its superior complementarity with existing approaches.
Furthermore, We further explore the influence of each component and observe that all components positively contribute to the performance of \toolname{}, \eg an accuracy improvement of 7.73\% brought by the hybrid retriever.
Finally, we implement \toolname{} with CodeBERT, GraphCodeBERT, UniXcoder and UniXcoder, and find these variants achieve an accuracy of 57.69\%$\sim$60.25\% and 46.96\%$\sim$48.58\% on \olddata{} and \newdata{}, outperforming the most recent approach \edit{} by 7.91\%$\sim$12.70\% and 5.86\%$\sim$9.51\%, respectively.
The results show that \toolname{} is universal and is effectively adapted to different PLMs with sustaining advancements, highlighting the applicability and generalizability of \toolname{} in practice.

To sum up, the contributions of this paper are as follows:
\begin{itemize}

  \item We introduce a generic yet effective retrieval-augmented assertion generation paradigm in the unit testing scenario with a hybrid assertion retriever and a PLM-based assertion generator.
  The framework is generic and can be integrated with different encoder-decoder PLMs.

  \item We propose \toolname{}, which utilizes a sparse token-based retriever and a dense embedding-based retriever to jointly search for the relevant assertion based on both lexical and semantic matching.
  We adopt a code-aware PLM CodeT5 as the foundation model for the assertion generator.
  To the best of our knowledge, \toolname{} is the first attempt to fine-tune recent PLMs with the advance of external codebases for the crucial assertion generation task.

  \item We extensively evaluate {\toolname} against six prior AG approaches on two datasets and two metrics.
  The experimental results demonstrate that {\toolname} significantly outperforms all baselines, with an average accuracy improvement of 58.81\% and 42.51\% on \olddata{} and \newdata{}.

  \item \revise{We release the relevant materials in our experiment to facilitate follow-up AG studies at the repository\footnote{\url{https://github.com/iSEngLab/RetriGen}}, including datasets, scripts, models, and generated assertions.}
  
\end{itemize}

\section{Background and Related Work}
\label{sec:background}
\subsection{Deep Assertion Generation}
\label{sec:background_AG}

With the success of deep learning (DL), researchers have increasingly been utilizing advanced DL techniques to automate a variety of software engineering tasks~\cite{watson2022systematic,yang2022survey,zhang2023survey}.
For example, Watson~\etal~\cite{watson2020learning} propose \atla{}, the first DL-based assertion generation approach that utilizes deep neural networks to learn correct assertions from existing test-assert pairs.
Yu~\etal~\cite{yu2022automated} propose two retrieval-based approaches to generate assertions by searching for similar assertions given a focal-test, namely IR-based assertion retrieval (\irar{}) and retrieved-assertion adaptation ($RA_{adapt}$ ).
\irar{} retrieves the assertion with the highest similarity to the given focal-test using measures like Jaccard similarity.
As \irar{} may not always retrieve completely accurate assertions, $RA_{adapt}$ attempts to revise the retrieved assertions by replacing tokens within them.
Two strategies are proposed for determining replacement values, \ie a heuristic-based approach \rah{} and a neural network-based approach \rann{}.
Furthermore, Yu~\etal~\cite{yu2022automated} propose an integration-based approach called \inte{} by integrating IR and DL techniques.
Building on \inte{}, Sun~\etal~\cite{sun2023revisiting} introduce \edit{} by searching for a similar focal-test, which is then modified by a neural model.
Unlike prior AG studies retrieving relevant assertions~\cite{yu2022automated} with token matching or training generator from scratch~\cite{watson2020learning} with historical test-assert pairs, our work attempts to utilize PLMs as an assertion generator with the help of a hybrid assertion retriever for more effective retrieval and generation.

The literature has also witnessed several studies~\cite{tufano2022generating, mastropaolo2021studying, mastropaolo2022using}, exploring the use of PLMs like T5~\cite{raffel2020exploring} in the field of assertion generation.
For instance, Mastropaolo~\etal~\cite{mastropaolo2022using} pre-train a T5 model and fine-tune it on four code-related tasks: bug-fixing, mutant generation, assertion generation, and code summarization.
These studies typically pre-train language models from scratch and fine-tune them on multiple downstream tasks.
In contrast, our work attempts to propose a specific AG approach by leveraging off-the-shelf PLMs. 
Nashid~\etal~\cite{nashid2023retrieval} propose CEDAR, a large language model (LLM)-based approach to apply retrieval-based demonstration selection strategy for program repair and assertion generation.
\toolname{} and CEDAR are fundamentally distinct regarding their learning paradigms, retrievers, and generators.
First, CEDAR utilizes few-shot learning with prompt engineering, while \toolname{} leverages fine-tuning with augmented inputs.
Second, CEDAR utilizes a single retriever, while \toolname{} builds a hybrid retriever to identify relevant assertions jointly, indicating superiority over a single retriever. 
Third, CEDAR queries APIs from a black-box LLM Codex with 12 billion parameters, whereas RetriGen fine-tunes an open-source PLM CodeT5 with only 220 million parameters. 
\revise{Recently, Zhang~\etal~\cite{zhang_exploring_2024} conduct an empirical study to explore the potential of several PLMs for generating assertions in a fine-tuning scenario.}

\subsection{Pre-trained Language Model}
\label{sec:background_LLM}

Recently, researchers have explored the capabilities of PLMs to revolutionize various code-related tasks~\cite{fan2023large,zhang2023survey_se}, such as code review~\cite{tufano2022using, li2022automating}, test generation~\cite{yuan2023no,xie2023chatunitest,wang2024hits,gu2024testart,tang2024chatgpt,yang2024evaluation,ryan2024code,lemieux2023codamosa} and program repair~\cite{zhang2023gamma,xia2023automated,zhang2024systematic}.

There exist three main categories based on model architectures.
(1) Encoder-only PLMs, \eg CodeBERT~\cite{feng2020codebert}, train the encoder part of the Transformer with masked language modeling, thus suitable for code understanding.
(2) Decoder-only PLMs, \eg CodeGPT~\cite{lu2021codexglue}, train the decoder part of the Transformer with unidirectional language modeling, thus suitable for auto-regressive generation.
(3) Encoder-decoder PLMs, \eg CodeT5~\cite{wang2021codet5}, train both encoder and decoder parts of the Transformer with denoising objectives, thus suitable for sequence-to-sequence generation.

In our work, the foundation model selection of \toolname{} is limited to encoder-decoder PLMs, as assertions are generated in a sequence-to-sequence learning manner.
Following prior work~\cite{wang2023rap,fu2022vulrepair,zhang2023pre,zhou2024out}, we consider CodeT5, a generic code-aware language model that is pre-trained on a large code corpus, achieving state-of-the-art performance in both code understanding and generation tasks.
CodeT5 is the most popular encoder-decoder PLM that is adopted by previous fine-tuning-based sequence-to-sequence code generation tasks~\cite{wang2024software,yuan2022circle,zhu2024grammart5}.
\revise{Besides, CodeT5 is trained with CodeSearchNet~\cite{lu2021codexglue} without test code snippets, thus avoiding the data leakage issue.}

\subsection{Information Retrieval for SE}
\label{sec:background_IR}

Information Retrieval (IR) refers to the process of identifying relevant information from a large collection of data.
IR has been extensively applied to a range of code-related tasks, such as fault localization~\cite{dao2017does} and test case prioritization~\cite{peng2020empirically}.
Such approaches search for the object that best matches the given query from the database based on different similarity coefficients, such as Jaccard similarity.
Besides, the literature has seen an increasing number of studies performing generation tasks with the retrieval-augmented paradigm, such as code repair~\cite{wang2023rap,nashid2023retrieval} and code summarization~\cite{li2021editsum,parvez2021retrieval}.
This paradigm improves the quality of generated results by grounding the model with external knowledge sources to supplement the PLM's internal representation of information~\cite{lewis2020retrieval}.
In our work, inspired by the intuition of retrieval-augmented generation in the PLM domain, we focus on the assertion generation problem and propose \toolname{} to fine-tune a PLM-based assertion generator with a hybrid assertion retriever.

\section{Approach}
\label{sec:approach}

\begin{figure*}
    \centering
    \includegraphics[width=0.85\linewidth]{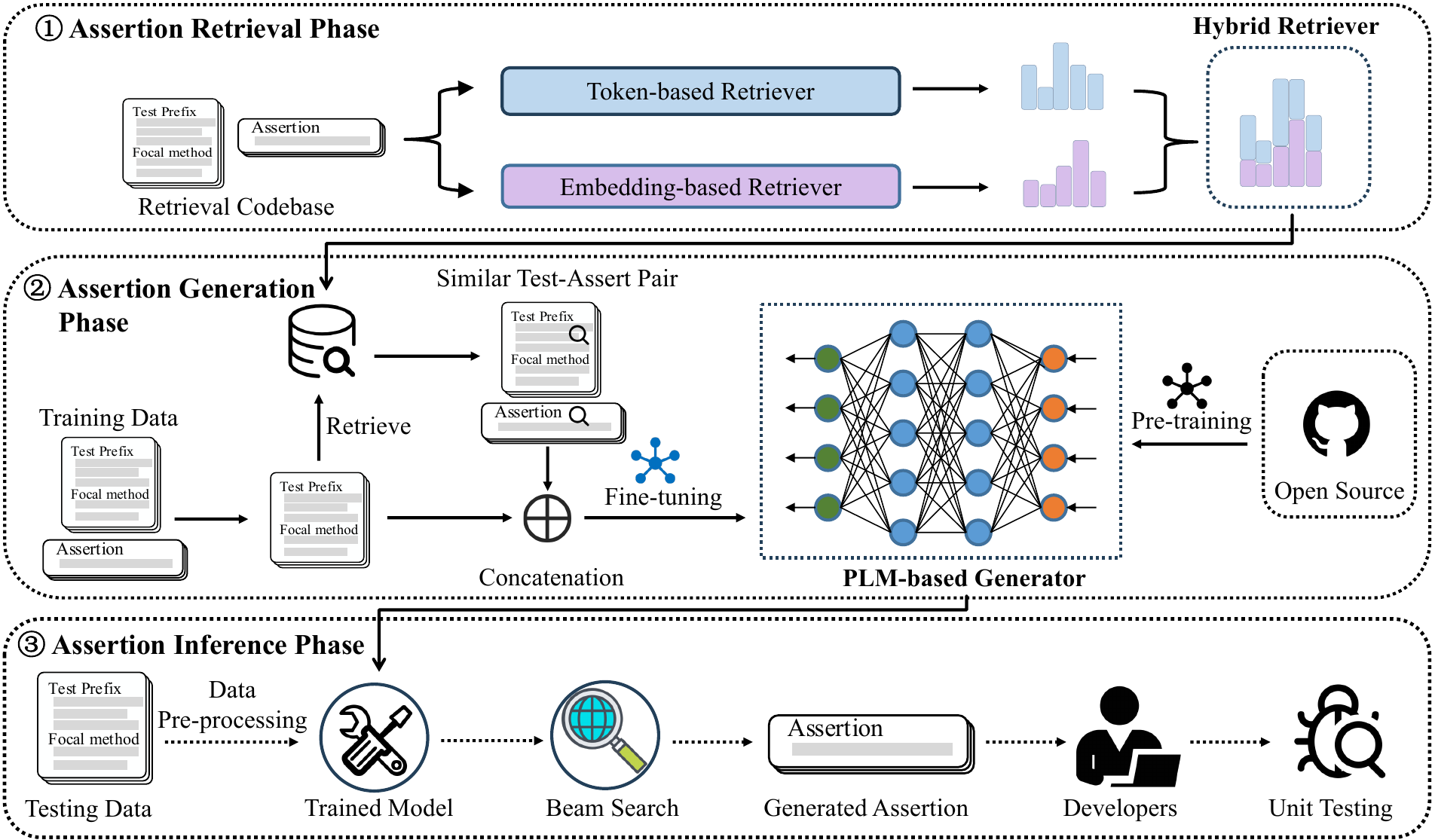}
    \Description{Framework Overview of \toolname{}}
    \caption{Framework Overview of \delete{our approach}\revise{\toolname{}}}
    \label{fig:workflow}
\end{figure*}

The framework overview of \toolname{} is illustrated in Fig.~\ref{fig:workflow}, which consists of three phases.
In the assertion retrieval phase, \toolname{} identifies a similar assertion from the external codebase based on lexical and semantical similarity calculated by a token-based retriever and an embedding-based retriever.
In the assertion generator training phase, 
\toolname{} is first pre-trained with millions of code snippets from open-source projects and then fine-tuned with the retrieval-augmented labeled pairs, \ie the focal-test and the retrieved assertion as input and the correct assertion as output.
In the assertion inference phase, after the generator is well trained, \toolname{} leverages the beam search strategy to generate a ranked list of candidate assertions, and return the one with the highest probability of being correct.

\subsection{Problem Definition}
\label{sec:approach_task_formulation}

Similar to the DL-based approach \atla{}, {\toolname} regards assertion generation as an NMT task based on the encoder-decoder Transformer architecture.
Suppose $\mathcal{D} = ({FT_i, A_i)}^{|\mathcal{D}|}_{i=1}$ be a unit testing dataset consisting of $|\mathcal{D}|$ test-assert pairs,
where $ FT_i$ and $A_i$ are the $i$-th focal-test and the corresponding assertion, respectively.
The sequence-to-sequence assertion generator attempts to predict $A_i$ from $FT_i$ in an auto-regressive manner, which is formally defined as follows:

\find{
\begin{definition}\textbf{Deep Assertion Generation.}\\
Given a focal-test input $FT_i=\left[ft_1, \cdots, f_m\right]$ with $m$ code tokens and an assertion output $A_i=\left[a_1, \ldots, a_n\right]$ with $n$ code tokens, 
the problem of assertion generation is defined as maximizing the conditional probability, \ie the likelihood of $A_i$ being the correct assertion:
\begin{displaymath}
P_{\theta}(A_i|FT_i)=\prod_{j=1}^{n}P_{\theta}
(a_j|a_1, \cdots, a_{j-1}; ft_1, \cdots, ft_m) 
\end{displaymath}
\label{def:generation}
\end{definition}
}

However, different from \atla{} generating new assertions from scratch, \toolname{} takes the focal-test and an additional retrieved assertion as input and the correct assertion as output.
Thus, on top of Definition~\ref{def:generation}, suppose $\mathcal{C} = {(FT'_j, A'_j)}^{|\mathcal{C}|}_{j=1}$ be an external codebase containing a large collection of historical test-assert pairs, where $FT'_j$ and $A'_j$ denote the $j$-th previous focal-test, and the corresponding assertion.
Then, with the codebase $\mathcal{C}$, the retrieval-augmented deep assertion generation can be defined as follows:

\find{
\begin{definition}\textbf{\small{Retrieval-Augmented Deep Assertion Generation.}}\\
Given a focal-test $FT_i$ in $\mathcal{D}$, the retriever searches for the most relevant focal-test $FT'_j$ from the codebase $\mathcal{C}$, and its assertion $A'_j=[a'_1, \cdots, a'_z]$ with $z$ code tokens.
Then the original focal-test input $FT_i $ is augmented with the retrieved assertion to form a new input sequence $\hat{FT_i}=FT_i\oplus A'_j $, where $\oplus $ denotes the concatenation operation.
Finally, the assertion generator attempts to generate $A_i$ from $\hat{FT_i}$ by learning the following the probability :
\small
\begin{displaymath}
P_{\theta}(A_i|\hat{FT_i})=\prod_{j=1}^{n}P_{\theta}
(a_j|a_1, \cdots, a_{j-1}; \underbrace{FT_i}_{\text{Original}}; \underbrace{a'_1, \cdots, a'_z }_{\text{Augmented}}) 
\end{displaymath}
\label{def:rag_generation}
\end{definition}
}

\subsection{Hybrid Assertion Retrieval Component}
\label{sec:approach_retriever}

\toolname{} utilizes a hybrid strategy that integrates a token-based retriever and an embedding-based retriever, allowing it to account for both the lexical and semantic similarities of test-assert pairs.

\subsubsection{Token-based Retriever}
\label{sec:approach__retriever_ir}
\toolname{} utilizes the sparse strategy IR as the token-based retriever to identify an assertion that closely resembles the query focal-test through lexical matching,  which has been proven to be effective in previous AG studies~\cite{yu2022automated,sun2023revisiting}.
In particular, \toolname{} tokenizes all focal-tests in both the dataset $\mathcal{D}$ and the codebase $\mathcal{C}$, and eliminates duplicate tokens to enhance the efficiency of the retrieval process.
Given a query focal-test $FT_i$ in $\mathcal{D}$, \toolname{} computes the lexical similarity between $FT_i$ and all focal-tests in $\mathcal{C}$ using the Jaccard coefficient as the similarity measure.
The Jaccard coefficient is a commonly employed metric for assessing the similarity between two sparse vector representations by considering their overlapping and unique tokens. 
The calculation of Jaccard similarity is defined in Formula~\ref{equation:jaccard}, where $S(FT_i)$ and $S(FT_j)$ represent the sets of code tokens for the two focal-tests $FT_i$ and $FT_j$, respectively.

\begin{equation}
    Jac(FT_i, FT_j) = \frac{|S(FT_i) \cap S(FT_j) |}{|S(FT_i) \cup S(FT_j) |}
\label{equation:jaccard}
\end{equation}

\subsubsection{Embedding-based Retriever}
\label{sec:approach_retriever_embedding}

\toolname{} utilizes the pre-trained CodeLlama~\cite{roziere2023code} as the embedding-based retriever, to identify the most similar assertion based on semantic similarity.
CodeLlama is an advanced language model pre-trained with common programming languages that support code understanding and generation tasks.
Unlike previous studies training the dense retriever from scratch~\cite{karpukhin2020dense}, we directly employ the checkpoint of CodeLlama from Hugging Face without any fine-tuning in our work as it is already trained with a mass of testing code to get meaningful vector embeddings for the query focal-test in the unit testing scenario. 

In particular, \toolname{} first splits the source code of the focal-test into a sequence of code tokens and utilizes CodeLlama to convert these tokenized elements into vector representations.
\toolname{} then computes the Cosine similarity between the embeddings of two focal-tests to assess their semantic relevance. 
Cosine similarity has been widely employed in prior research for measuring the semantic relationship between two dense vectors~\cite{pan2023ltm}.
It is calculated by taking the cosine of the angle between two vectors, which is the dot product of the vectors divided by the product of their magnitudes.
Formula~\ref{equation:cosine} presents the Cosine similarity calculation, where $\mathbf{FT_i}$ and $\mathbf{FT_j}$ represent the embeddings of focal-tests $FT_i$ and $FT_j$.

\begin{equation}
    Cos(FT_i, FT_j) = \frac{\mathbf{FT_i} \cdot \mathbf{FT_j}}{\| \mathbf{FT_i}\|  \|\mathbf{FT_j} \|}
\label{equation:cosine}
\end{equation}

\subsubsection{Hybrid Retriever}
\label{sec:appraoch_retriever_hybrid}
Our sparse token-based retriever primarily relies on identifier naming in the source code, making it more sensitive to lexical choices rather than the underlying code semantics.
In contrast, our dense embedding-based retriever focuses more on the semantic representation of the code rather than its literal syntax.
Thus, to incorporate both lexical and semantic information, we adopt a hybrid approach that integrates the two retrievers in a unified manner. 
The combined similarity score between two focal-tests is computed as shown in Formula~\ref{equation:hybrid}, where $\lambda$ serves as a weighting factor to balance the contributions of the two retrievers, with its default value set to 1.
Based on this combined similarity score, we select the top-1 relevant test-assert pair $(FT'_i, A'_i) $  to guide the assertion generator for generating assertions.

\begin{equation}
    Sim(FT_i, FTj)=Jac(FT_i,FTj) + \lambda Cos(FT_i, FTj)
\label{equation:hybrid}
\end{equation}

\subsection{Assertion Generation Component}
\label{sec:approach_generation}

\toolname{} utilizes CodeT5 as the foundation model to generate the assertion $A_i$, leveraging both the focal-test $FT_i$ and the externally retrieved assertion $A'_j$ as input.
The assertion generation component is general to a mass of existing PLMs, and we select CodeT5 as it is a code-aware encoder-decoder PLM optimized for source code and has demonstrated effectiveness in various sequence-to-sequence code generation tasks, such as program repair~\cite{fu2022vulrepair,zhang2023pre}.

\textbf{Input Representation.}
As discussed in Section~\ref{sec:approach_task_formulation}, the retrieval-augmented focal-test input to \toolname{} is represented as $\hat{FT_i}=FT_i \oplus A'_j$ with the concatenation operation $\oplus$.
Thus, the generator is trained to learn the transformation patterns from the retrieval-augmented input $\hat{FT_i}$ to the output $A_i$ by the sequence-to-sequence learning.

\textbf{Model Architecture.}
The generation model of \toolname{} is composed of an encoder stack and a decoder stack, culminating in a linear layer with softmax activation.
First, \toolname{} splits the source code of the input $\hat{FT_i}$ into subwords using a code-specific Byte-Pair Encoding (BPE) tokenizer, which has been pre-trained on eight widely used programming languages and is specifically designed for tokenizing source code~\cite{wang2021codet5}.
The BPE tokenizer is able to break down rare tokens into meaningful subword units based on their frequency distribution to address the common Out-Of-Vocabulary (OOV) problem in code-related tasks.
Second, \toolname{} employs a word embedding stack to generate representation vectors for tokenized focal-test functions, allowing it to capture the semantic meaning of the code tokens as well as their positional relationships within the code snippet.
Third, \toolname{} inputs the vectors into an encoder stack to obtain the hidden state, which is subsequently passed to a decoder stack for further processing.
Finally, the output from the decoder stack is passed through a linear layer with softmax activation, generating a probability distribution over the vocabulary.

\textbf{Loss Function.}
The assertion generation model receives $\hat{FT_i}$ as input and produces the correct assertion $A_i$ by learning the mapping between $\hat{FT_i}$ and $A_i$.
Thus, the model's parameters are adjusted through the training dataset, with the goal of improving the mapping by maximizing the conditional probability $P_{\theta}(A_i|\hat{FT_i})$.
\toolname{} employs the common cross entropy as the loss function to train the assertion generator $\mathcal{L}$, which is widely adopted in previous code-related studies~\cite{wang2021codet5, feng2020codebert}.
The cross-entropy loss is minimized between each position in the predicted assertion and each position
in the ground-truth assertion, defined as follows.

\begin{equation}
    \mathcal{L} = -\sum_{i=1}^{|\mathcal{D}|}\log(P_{\theta}(A_i|\hat{FT_i}))
\label{equation:loss}
\end{equation}

\subsection{Assertion Inference}
During the model inference phase, once the generation model is effectively trained, given a focal-test as the input, \toolname{} employs the beam search strategy to produce a ranked list of assertion candidates based on the vocabulary's probability distribution.
In particular, beam search~\cite{watson2020learning} is a widely used strategy for identifying the highest-scoring candidate assertions by progressively prioritizing the top-$k$ probable tokens according to their predicted likelihood scores.
The correctness of the generated candidate assertion can be automatically evaluated by comparing it with ground-truth assertions or manually inspected by test experts for deployment in the unit testing pipeline.

\section{Experimental Setup}
\label{sec:experimental_setup}

\subsection{Research Questions}
To evaluate the performance of \toolname{}, we address the following three research questions (RQs):

\begin{itemize}
    \item RQ1: How does \toolname{} compare to state-of-the-art assertion generation approaches?
    \item RQ2: To what extent do different choices influence the overall effectiveness of \toolname{}?
    \item RQ3: What is the generalizability of \toolname{} when utilizing other advanced PLMs?
\end{itemize}

\begin{table*}[t]
  \centering
  \caption{Detailed statistics of each type in $Data_{new}$ and $Data_{old}$}
  \resizebox{\linewidth}{!}{
    \begin{tabular}{c|c|ccccccccc}
    \toprule
    \textbf{AssertType} & Total & Equals & True  & That  & NotNull & False & Null  & ArrayEquals & Same  & Other \\ 
    \midrule
    $Data_{old}$ & 15,676 & 7,866 (50\%) & 2,783 (18\%) & 1,441 (9\%) & 1,162 (7\%) & 1,006 (6\%) & 798 (5\%) & 307 (2\%) & 311 (2\%) & 2 (0\%) \\
    $Data_{new}$ & 26,542 & 12,557 (47\%) & 3,652 (14\%) & 3,532 (13\%) & 1,284 (5\%) & 1,071 (4\%) & 735 (3\%) & 362 (1\%) & 319 (1\%) & 3,030 (11\%) \\
    \bottomrule
    \end{tabular}}
  \label{tab:statistics}
\end{table*}

\subsection{Datasets}
\label{sec:dataset}

Following \edit{}~\cite{sun2023revisiting}, we utilize two popular large-scale datasets, \ie \olddata{} and \newdata{}, to evaluate \toolname{} and the baselines. 
These datasets have been widely used in previous deep assertion generation studies~\cite{watson2020learning, yu2022automated, sun2023revisiting, he2024empirical}, including all of our baselines. 
A brief introduction to both datasets is provided below.

(1) \olddata{}. 
$Data_{old}$ is initially constructed by Watson~\etal~\cite{watson2020learning} to propose \atla{}~\cite{watson2020learning}, the first deep assertion generation approach.
Watson~\etal~\cite{watson2020learning} collect a dataset of 2.5 million test methods within GitHub, each comprising test prefixes and their corresponding assertion statements.
In $Data_{old}$, every test method is linked to a specific focal method, which represents the production code being tested.
Watson~\etal then perform preprocessing to remove test methods with token lengths exceeding 1K and to filter out assertions that contain \emph{unknown} tokens not present in the focal test or the vocabulary.

(2) \newdata{}.
However, \olddata{} excludes assertions containing unknown tokens to oversimplify the assertion generation problem, thus being unsuitable to reflect the real-world data distribution.
Thus, Yu~\etal~\cite{yu2022automated} address this issue by creating an extended dataset, referred to as $Data_{new}$, which includes additional 108,660 samples that are excluded due to unknown tokens in \olddata{}.

As a result, \olddata{} and \newdata{} contain a total of 156,760 and 265,420 samples, respectively. 
These datasets are divided into training, validation, and test sets using an 8:1:1 ratio, as done by Watson~\etal~\cite{watson2020learning} and Yu~\etal~\cite{yu2022automated}. 
In this paper, we strictly adhere to the replication package provided by prior work~\cite{watson2020learning,yu2022automated, sun2023revisiting}. 
Table~\ref{tab:statistics} presents the statistics of the testing sets for both datasets and their distribution across various assertion types.

\subsection{Baselines}
\label{sec:baselines}
To address the above-mentioned RQs, we compare {\toolname} against six state-of-the-art AG approaches, including DL-based, retrieval-based, and integration-based ones.
First, we include \atla{}, the first and classical AG technique that utilizes a sequence-to-sequence model to generate assertions from scratch. 
\atla{} is the most relevant baseline as both \atla{} and \toolname{} consider assertion generation as an NMT task based on the encoder-decoder Transformer architecture. 
Second, we incorporate three existing retrieval-based AG techniques: \irar{}, \rah{}, and \rann{}.
Finally, we consider one state-of-the-art integration-based approach \inte{}, and its most recent follow-up \edit{}, as detailed in Section~\ref{sec:background_AG}.
It is worth noting that we exclude CEDAR as a baseline mainly due to the severe data leakage issue of the utilized black-box LLM Codex.
LLMs are proprietary, and their training details (\eg pre-training datasets) are not publicly disclosed, making it difficult to ensure whether models have been exposed to the evaluation dataset during training~\cite{tian2024debugbench,silva2024gitbug,zhang2023critical}.

\subsection{Evaluation Metrics}
\label{sec:metrics}

Following prior AG work~\cite{watson2020learning,yu2022automated,sun2023revisiting}, we consider two metrics to evaluate the performance of \toolname{} and baselines, which are widely-adopted in code-related studies~\cite{wang2023rap,tufano2022using,zhang2023pre}.

\textbf{\emph{Accuracy.}} 
Accuracy is defined as the proportion of focal tests for which assertions generated by AG approaches match the ground-truth ones, and is utilized by all baseline methods~\cite{watson2020learning,yu2022automated,sun2023revisiting}.
An assertion is deemed correct only when it precisely aligns with the ground truth at the token level.

\textbf{\emph{CodeBLEU.}}
Apart from accuracy, BLEU is utilized by previous AG studies~\cite{watson2020learning,yu2022automated} to evaluate how closely the generated assertion matches the ground truth.
However, BLEU is initially developed for natural language through token-level matching, overlooking significant lexical and semantic aspects of source code.
In this work, we consider a code-aware variant of BLEU, \ie CodeBLEU, which is designed particularly for code generation tasks~\cite{lu2021codexglue}.
Compared with BLEU, CodeBLEU is more suitable for the AG task as it further incorporates lexical similarity with the AST information and semantic similarity with a data-flow structure.

\subsection{Implementation Details}

We implement our experiments, including \toolname{}, using the PyTorch framework \cite{PyTorch}. 
We employ the Hugging Face implementation \cite{huggingface} of the studied PLMs and all hyperparameters are taken from default values.
We utilize the training set as the search codebase, consistent with prior deep AG studies~\cite{yu2022automated,sun2023revisiting}.
To prevent data leakage, we perform queries on the datasets to verify that there is no overlap between the evaluation and training sets.
This strategy ensures that the retriever and generator are not privy to ground-truth assertions.
\revise{During training, we set the batch size to 8, the maximum lengths of input to 512, the maximum lengths of output to 256, and the learning rate to 2e-5.}
All training and evaluations are performed on one Ubuntu 20.04 server with two NVIDIA GeForce RTX 4090 GPUs.

\section{Evaluation and Results}
\label{sec:results}

\subsection{RQ1: Comparison with State-of-the-arts}

\textbf{\emph{Experimental Design.}} 
In RQ1, we aim to evaluate the effectiveness of assertions generated by {\toolname}.
We compare {\toolname} with six state-of-the-art AG techniques, including \atla{}, \irar{}, \rah{}, \rann{}, \inte{} and \edit{}.
To ensure a fair comparison, we employ the same training set to train baselines and \toolname{}, and we provide the same evaluation set to all approaches for evaluation.
We calculate the accuracy and CodeBLEU of all baselines and \toolname{} by comparing generated assertions and human-written assertions.
\revise{According to prior work~\cite{sun2023revisiting, yu2022automated}, we directly utilize reported results from their original paper rather than re-executing baselines, thereby minimizing potential risks.
}

\begin{table*}[t]
\footnotesize
  \centering
  \caption{Comparisons of \toolname{} with state-of-the-art AG approaches}
    \begin{tabular}{c|cc|cc}

    \toprule
    \multirow{2}[4]{*}{\textbf{Appraoch}} & \multicolumn{2}{c|}{\olddata{}} & \multicolumn{2}{c}{\newdata{}} \\
\cmidrule{2-5}          & \textbf{Accuracy} & \textbf{CodeBLEU} & \multicolumn{1}{c}{\textbf{Accuracy}} & \textbf{CodeBLEU} \\
    \midrule
    \atla{} & 31.42\% ($\uparrow$103.98\%) & 63.60\% ($\uparrow$25.46\%) & 21.66\% ($\uparrow$136.47\%) & 37.91\% ($\uparrow$75.89\%) \\
    \irar{} & 36.26\% ($\uparrow$76.75\%) & 71.03\% ($\uparrow$12.33\%) & 37.90\% ($\uparrow$35.15\%) & 62.67\% ($\uparrow$6.40\%) \\
    \rah{} & 40.97\% ($\uparrow$56.43\%) & 72.46\% ($\uparrow$10.12\%) & 39.65\% ($\uparrow$29.18\%) & 63.66\% ($\uparrow$4.74\%) \\
    \rann{} & 43.63\% ($\uparrow$58.13\%) & 72.12\% ($\uparrow$10.64\%) & 40.53\% ($\uparrow$17.40\%) & 63.19\% ($\uparrow$5.52\%) \\
    \inte{} & 46.54\% ($\uparrow$37.71\%) & 73.29\% ($\uparrow$8.87\%) & 42.20\% ($\uparrow$21.37\%) & 63.00\% ($\uparrow$5.84\%) \\
    \edit{} & 53.46\% ($\uparrow$19.88\%) & 77.00\% ($\uparrow$3.62\%) & 44.36\% ($\uparrow$15.46\%) & 64.40\% ($\uparrow$3.54\%) \\
    \midrule
    \toolname{} & \textbf{64.09\%} & \textbf{79.79\%} & \multicolumn{1}{c}{\textbf{51.22\%}} & \textbf{66.68\%} \\
    \bottomrule
      
     \multicolumn{5}{l}{$\uparrow$ denotes performance improvement of \toolname{} against state-of-the-art baselines}
    \end{tabular}
  \label{tab:rq1}
\end{table*}%

\textbf{\emph{Results and Analysis.}}
Table~\ref{tab:rq1} presents the comparison results of \toolname{} and six baselines in terms of the prediction accuracy and CodeBLEU score.
Overall, {\toolname} achieves a remarkable performance with a prediction accuracy of 57.66\% and a CodeBLEU score of 73.24\% on average for two datasets, with an improvement of 50.66\% and 14.14\% against all previous baselines, respectively.

In particular, when compared with the DL-based approach \atla{}, {\toolname} achieves a prediction accuracy of 64.09\% and 51.22\% on the \olddata{} and \newdata{} datasets, yielding a remarkable improvement of 103.98\% and 136.47\%, respectively.
Similarly, \toolname{} exhibits substantial gains in the CodeBLEU metric, achieving improvements of 25.46\% and 75.89\% on average.
It is worth noting that \atla{} is the most relevant to our \toolname{}, as both approaches address assertion generation as an NMT task based on the encoder-decoder Transformer architecture.
The significant improvement against \atla{} indicates the advantage of our PLM-based assertion generator, which is pre-trained with millions of code snippets, thus getting rid of the limited number of training samples.
When compared with retrieval-based approaches, {\toolname} attains an average accuracy improvement of 55.95\%, 42.81\%, and 37.76\% against \irar{}, \rah{}, and \rann{} across the two datasets.
Similarly, in terms of the CodeBLEU metric,  the improvement still reaches 9.37\%, 7.43\%, and 8.08\%.
When compared with the two advanced integration-based approaches \inte{} and \edit{}, \toolname{} improves the prediction accuracy by 37.71\% and 19.88\% on \olddata{}, and 21.37\% and 15.46\% on \newdata{}, respectively.

\begin{table*}[t]
  \centering
  \caption{Detailed statistics of \toolname{} and baselines for each assert type}
  \resizebox{0.98\linewidth}{!}{
    \begin{tabular}{c|c|c|ccccccccc}
    \toprule
    \multirow{2}{*}{\textbf{Dataset}} & \multirow{2}{*}{\textbf{Approach}} & \multirow{2}{*}{\textbf{Total}} & \multicolumn{9}{c}{\textbf{AssertType}} \\ \cmidrule{4-12}
	&       &       & \textbf{Equals} & \textbf{True} & \textbf{That} & \textbf{NotNull} & \textbf{False} & \textbf{Null} & \textbf{ArrayEquals} & \textbf{Same} & \textbf{Other} \\
    \midrule
    \multirow{7}{*}{\olddata{}} & \atla{} & 4,925 (31\%) & 2,501 (32\%) & 966 (35\%) & 248 (17\%) & 598 (51\%) & 229 (23\%) & 236 (30\%) & 100 (33\%) & 47 (15\%) & 0 (0\%) \\
          & \irar{}    & 5,684 (36\%) & 2,957 (38\%) & 1,039 (37\%) & 449 (31\%) & 439 (38\%) & 314 (31\%) & 285 (36\%) & 111 (36\%) & 89 (29\%) & \textbf{1 (50\%)} \\
          & \rah{}   & 6,423 (41\%)   & 3,300 (42\%)   & 1,151 (41\%)      & 536 (37\%)      & 553 (48\%)      &  335 (33\%)     &  316 (40\%)      & 120 (39\%)      &  111 (36\%)     & \textbf{1 (50\%)} \\
          & \rann{}   & 6,839 (44\%)      & 3,509 (45\%)      & 1,225 (44\%)      & 551 (38\%)      & 610 (52\%)      & 342 (34\%)      & 341 (43\%)       & 134 (44\%)      & 126 (41\%)      & \textbf{1 (50\%)} \\
          & \inte{} & 7,295 (47\%)      & 3,714 (47\%)      & 1,333 (48\%)      & 546 (38\%)      & 724 (62\%)      & 348 (35\%)      & 352 (44\%)       & 148 (48\%)       & 129 (41\%)      & \textbf{1 (50\%)}  \\
          & \edit{} & 8,380 (53\%)       & 4,131 (53\%)      & 1,581 (57\%)      & 526 (36\%)     & 807 (69\%)   & 577 (57\%)     & 469 (59\%)      & 167 (54\%)       & 122 (39\%)       & 0 (0\%) \\
          & {\toolname} & \textbf{10042 (64\%)}      &  \textbf{5027 (64\%)}   &  \textbf{1791 (64\%)}   & \textbf{804 (56\%)}     & \textbf{820 (71\%)}  & \textbf{654 (65\%)}    & \textbf{568 (71\%)}     & \textbf{184 (60\%) }  & \textbf{194 (62\%)}     & 0(0\%)    \\
    \midrule
    \multirow{7}{*}{\newdata{}} & \atla{} & 5,749 (22\%) & 2,900 (23\%) & 619 (17\%) & 537 (15\%) & 388 (30\%) & 126 (12\%) & 85 (12\%) & 47 (13\%) & 37 (12\%) & 1,010 (33\%) \\
          & \irar{}    & 10,059 (38\%) & 4,664 (37\%) & 1,436 (39\%) & 1,070 (30\%) & 600 (47\%) & 394 (37\%) & 286 (39\%) & 147 (41\%) & 113 (35\%) & 1,349 (45\%) \\
          & \rah{}   & 10,525 (40\%)      & 4,882 (39\%)      & 1,487 (41\%)       & 1,142 (32\%)      & 651 (51\%)      & 403 (38\%)      & 297 (40\%)      & 154 (43\%)       & 121 (38\%)       & 1,388 (46\%) \\
          & \rann{}   & 10,758 (41\%)       & 4,988 (40\%)      & 1,526 (42\%)      & 1,161 (33\%)      & 691 (54\%)       & 401 (37\%)      & 308 (42\%)      & 162 (45\%)      & 126 (39\%)       & 1,395 (46\%)  \\
          & \inte{} & 11,201 (42\%)      & 5,248 (42\%)      & 1,566 (43\%)       & 1,196 (34\%)       & 711 (55\%)       & 401 (37\%)      & 313 (43\%)      & 162 (45\%)      & 128 (40\%)      & 1,476 (49\%)  \\
          & \edit{} & 11,773 (44\%)       & 5,339 (42\%)      & 1,702 (47\%)       & 1,304 (37\%)       & 800 (62\%)       & 523 (49\%)      & 376 (51\%)       & 172 (47\%)       & 139 (44\%)      & 1,418 (47\%)   \\
           & {\toolname} & \textbf{13590 (51\%)}      & \textbf{6294 (50\%})  & \textbf{1860 (51\%)}    & \textbf{1588 (45\%)}    & \textbf{840 (65\%)}   & \textbf{609 (57\%)}    & \textbf{433 (59\%) }    & \textbf{195 (54\%)}      & \textbf{176 (55\%)}     & \textbf{1595 (53\%)}    \\
    \bottomrule
    \end{tabular}%
    }
  \label{tab:type}%
\end{table*}%

\textbf{\emph{Effectiveness on Different Assertion Types.}} 
Table~\ref{tab:type} presents the performance of {\toolname} and baselines on different types of assertions.
The rows denote seven approaches (\ie six baselines and \toolname{}) and two datasets (\ie \olddata{} and \newdata{}).
The columns denote nine different assertion types, with each cell displaying the number of correctly generated assertions and their corresponding ratios in parentheses.
From Table~\ref{tab:type}, \toolname{} outperforms all baselines on all standard JUnit assertion types across both datasets.
For example, for the most common \textit{Equals} type, \toolname{} correctly generates 5027 and 6294 assertions on two datasets, with a prediction accuracy rate of 64\% and 50\%, improving the best-performing one \edit{} by 20.75\% and 19.05\%, respectively.
Besides, in the case of \textit{Other} assertion type, \toolname{} outperforms existing baselines, except in the \olddata{} dataset, which has only two samples and is too small to yield convincing results.
For example, \toolname{} generates 1595 correct assertions on \newdata{} with an improvement of 8.16\% against the best-performing one \inte{}, indicating the capability of \toolname{} in addressing such non-standard JUnit assertion types.
In summary, the experimental results demonstrate the generality of \toolname{} in generating various types of assertions, including both standard and non-standard Junit types.

\begin{figure}[t]
\centering
    \subfigure[\olddata{}] {
        \includegraphics[width=0.4\columnwidth]{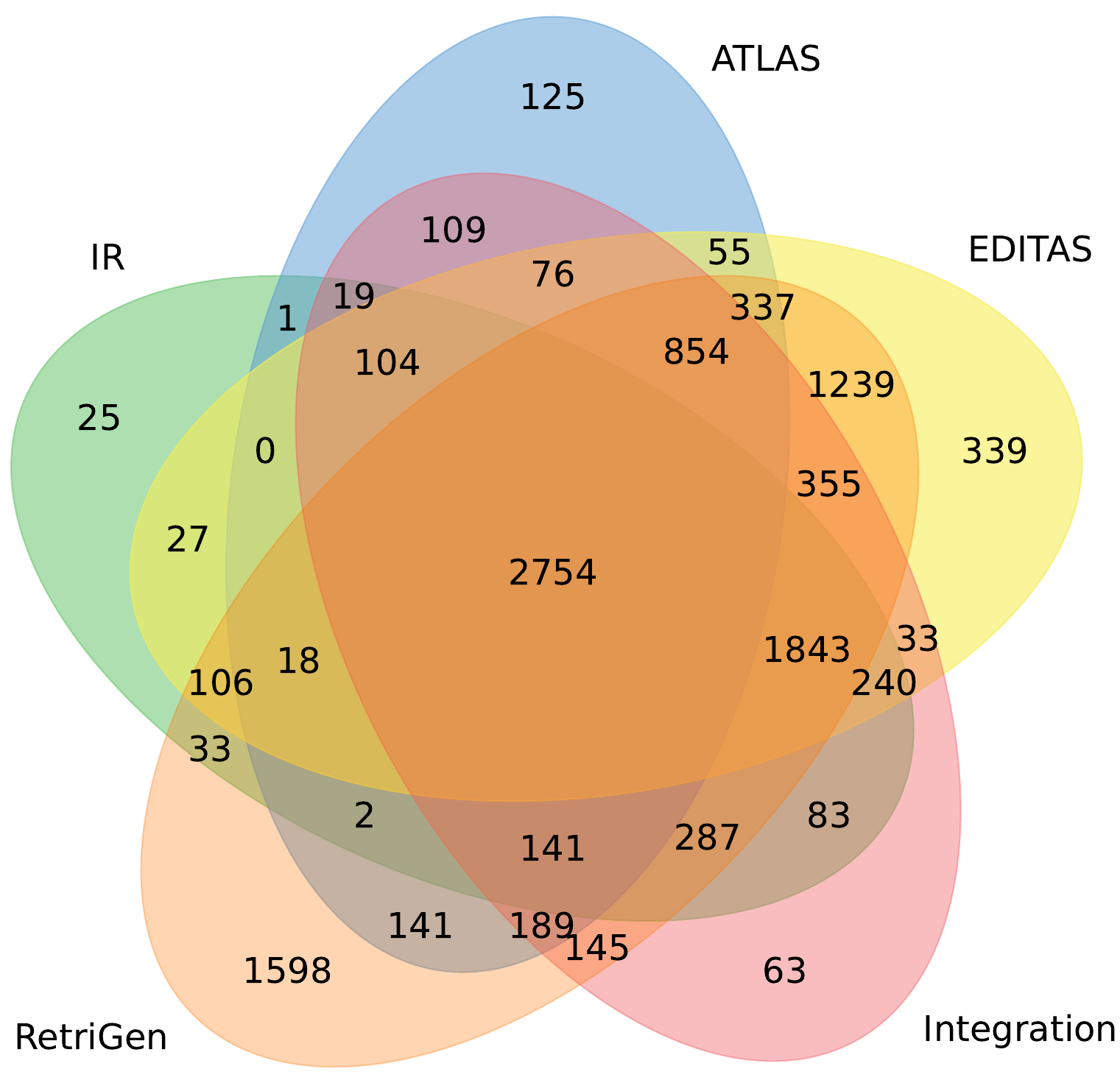}
        \label{fig:veen_old}
    }
    \subfigure[\newdata{}] {
        \includegraphics[width=0.4\columnwidth]{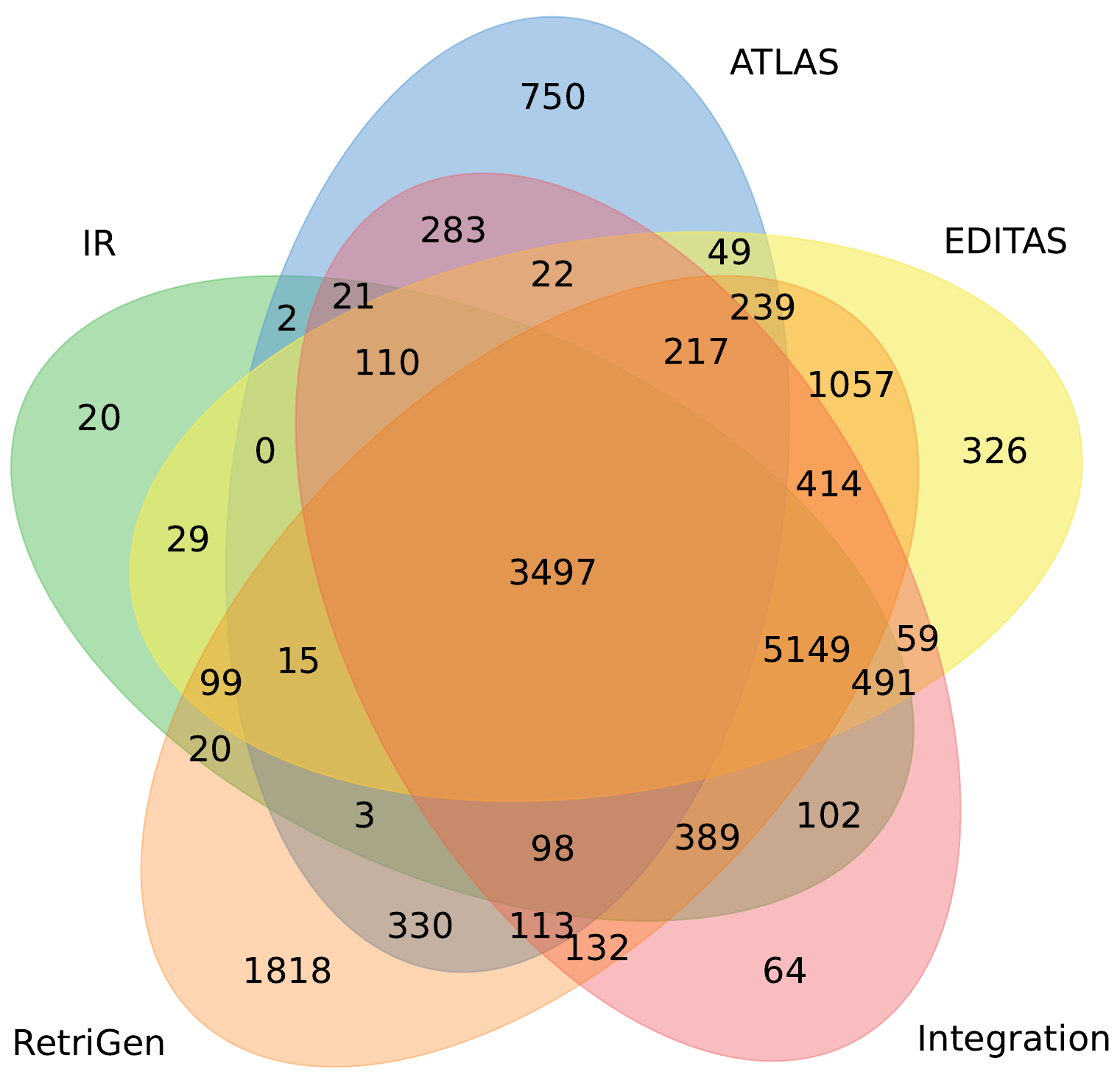}
        \label{fig:veen_new}
    }
\caption{Overlaps of assertions generated by different AG approaches}
\Description{Overlaps of assertions generated by different AG approaches}
\label{fig:veen}
\end{figure}

\textbf{\emph{Overlap Analysis.}}
To explore the extent to which \toolname{} complements prior studies, Fig.~\ref{fig:veen} presents the number of overlapping assertions generated by different AG techniques.
Due to page limit, we select one DL-based technique \atla{}, one retrieval-based technique \irar{}, and one integration-based technique \inte{}, as well as the most recent \edit{} for comparison.
As shown in Fig.~\ref{fig:veen}, {\toolname} generates 1598 unique assertions that other AG approaches fail to generate on \olddata{}, which are 1473, 1573, 1535, 1259 more than \atla{}, \irar{}, \inte{}, and \edit{}, respectively.
The improvement achieves substantial multiples of 11.78X, 62.92X, 24.37X, and 3.71X over these respective baselines.
Similarly, on the \newdata{} dataset, {\toolname} showcases its effectiveness by generating 1818 unique assertions, which is 1528 more than the baselines on average, with a remarkable 30.38X improvement. 
Overall, the results demonstrate the superior capability of {\toolname} in generating unique assertions, indicating the potential to complement existing AG techniques.

\finding{1}{Our comparison results demonstrate that: 
(1) {\toolname} significantly outperforms all baselines in terms of accuracy and CodeBLEU across both datasets, \eg with an average accuracy improvement of 58.81\% on \olddata{} and 42.51\% on \newdata{};
(2) \toolname{} consistently outperforms all baselines on all standard JUnit assertion types across both datasets, \eg improving \edit{} by 20.75\% and 19.05\% in generating assertions for the \textit{Equals} type;
(3) \toolname{} generates a large number of unique assertions that all baselines fail to generate across both datasets, \eg 1598 and 1818 unique ones on \olddata and \newdata{}, with an improvement of 3.71X and 4.58X against the most recent baseline \atla{}.
}

\subsection{RQ2: Impact Analysis}

\textbf{\emph{Experimental Design.}}
In RQ2, we attempt to investigate the impacts of different components in \toolname{}.
\toolname{} employs a hybrid retriever to search for a relevant test-assert pair, which consists of a token-based retriever and an embedding-based retriever.
To illustrate the importance of each component, we compare \toolname{} with three of its variants:
(1) \toolname{}$_\text{none}$ that generates assertions without any retriever, which is similar to \atla{}; 
(2) \toolname{}$_\text{token}$ that generates assertions with only a token-retriever;
(3) \toolname{}$_\text{embed}$ that generates assertions with only an embedding-retriever.

\begin{table*}[t]
\footnotesize
    \centering
    \caption{Effectiveness of {\toolname} with different retriever choices}

    \begin{tabular}{c|cc|c|c}
    \toprule
    \multirow{2}[4]{*}{\textbf{Appraoch}} & \multicolumn{2}{c|}{\olddata{}} & \multicolumn{2}{c}{\newdata{}} \\
\cmidrule{2-5}          & \textbf{Accuracy} & \textbf{CodeBLEU} & \multicolumn{1}{c}{\textbf{Accuracy}} & \textbf{CodeBLEU} \\
    \midrule
    \toolname{}$_\text{none}$  & 58.71\% ($\uparrow$9.16\%) & 74.91\% ($\uparrow$6.51\%) & 48.19\% ($\uparrow$6.29\%) & 63.40\% ($\uparrow$5.17\%) \\
    \toolname{}$_\text{token}$ & 63.37\% ($\uparrow$1.14\%) & 78.54\% ($\uparrow$1.59\%) & 51.15\% ($\uparrow$0.14\%) & 66.54\% ($\uparrow$0.21\%) \\
    \toolname{}$_\text{embed}$ & 63.11\% ($\uparrow$1.55\%) & 79.01\% ($\uparrow$0.99\%) & 50.74\% ($\uparrow$0.95\%) & 66.09\% ($\uparrow$0.89\%) \\
    \midrule
    \toolname{} & \textbf{64.09\%} & \textbf{79.79\%} & \multicolumn{1}{c}{\textbf{51.22\%}} & \textbf{66.68\%} \\
    \bottomrule
    \end{tabular}%

    \label{tab:rq2}
\end{table*}

\textbf{\emph{Results and Analysis.}}
Table~\ref{tab:rq2} displays the comparison results of \toolname{} under different components.
First, we find that \toolname{} outperforms all variants in terms of accuracy and CodeBLEU across both datasets, indicating the benefits of each component.
For example, on the \olddata{} dataset, the hybrid retriever of \toolname{} improves the prediction accuracy by 9.16\%, 1.14\% and 1.55\%, against \toolname{}$_\text{none}$, \toolname{}$_\text{token}$ and \toolname{}$_\text{embed}$, respectively.
Similarly, the average improvement reaches 2.46\% 3.03\%, and 2.09\% under the remaining three settings, \ie accuracy on \olddata{}{}, CodeBLEU on \olddata{} and \newdata{}.
Second, we find that without any retriever, \toolname{} still achieves remarkable performance, with a prediction accuracy of 58.71\% and 48.19\% on two datasets.
Compared with the most relevant baseline \atla{}, the improvement reaches 86.86\% and 122.48\%, highlighting \toolname{}'s advanced capability in overcoming the challenges posed by limited training datasets.
It is worth noting that although the improvement of \toolname{} over \toolname{}$_\text{token}$ and \toolname{}$_\text{embed}$ is not as significant as its improvement over \toolname{}$_\text{none}$, given that three variants have achieved state-of-the-art results, the improvement brought by our hybrid retriever is valuable in practice, leading to a new higher baseline of deep assertion generation performance.

\finding{2}{Our impact analysis indicates that 
all components (such as token-based and embedding-based retrievers) positively contribute to the effectiveness of \toolname{} across two metrics, leading to new records on two widely adopted datasets.
}

\subsection{RQ3: Generalizability of {\toolname}}
\label{sec:rq3}

\textbf{\emph{Experimental Design.}}
RQ1 and RQ2 have proven that \toolname{} outperforms the baselines when utilizing CodeT5 as the foundational model.
To further explore the impact of various PLMs on the performance of \toolname{}, we utilize four additional advanced PLMs for the assertion generation task: CodeBERT, GraphCodeBERT, UniXcoder, and CodeGPT.
Particularly, CodeBERT~\cite{feng2020codebert} is a bi-modal PLM for both programming language and natural language understanding, pre-trained with masked language modeling and replaced token detection.
GraphCodeBERT~\cite{guo2020graphcodebert} is a successor of CodeBERT by incorporating two structure-aware pre-training tasks, \ie edge prediction and node alignment.
UniXcoder~\cite{guo2022unixcoder} is a unified cross-modal PLM designed to enhance both code understanding and generation capabilities with two pre-training tasks, \ie multi-modal contrastive learning, and cross-modal generation.
All these PLMs are pre-trained with source code, publicly accessible, and medium-scale, thus suitable for fine-tuning in the AG task.

To implement \toolname{}, for encoder-decoder PLMs, like UniXcoder, we use Test-Assertion Pairs to train both encoder and decoder components, thus learning hidden mappings between two sequences.
For encoder-only PLMs, like CodeBERT and GraphCodeBERT, we initialize a new decoder from scratch to construct an encoder-decoder architecture for fine-tuning. 
For decoder-only PLMs, like CodeGPT, we directly leverage GPT's capabilities to generate subsequent assertions from previous focal-methods in an auto-regressive manner.

\begin{table*}[t]
\footnotesize
    \centering
    \caption{Effectiveness of {\toolname} with different PLMs as assertion generators}
    \begin{tabular}{c|cc|c|c}
    \toprule
    \multirow{2}[4]{*}{\textbf{Appraoch}} & \multicolumn{2}{c|}{\olddata{}} & \multicolumn{2}{c}{\newdata{}} \\
\cmidrule{2-5}          & \textbf{Accuracy} & \textbf{CodeBLEU} & \multicolumn{1}{c}{\textbf{Accuracy}} & \textbf{CodeBLEU} \\
    \midrule
    GraphCodebert  & 60.25\% ($\uparrow$6.37\%) & 77.42\% ($\uparrow$3.06\%) & 46.96\% ($\uparrow$9.07\%) & 63.05\% ($\uparrow$5.76\%) \\
    Unixcoder  & 59.17\% ($\uparrow$8.32\%) & 75.77\% ($\uparrow$5.31\%) & 48.58\% ($\uparrow$5.43\%) & 63.67\% ($\uparrow$4.73\%) \\
    CodeBERT  & 58.84\% ($\uparrow$8.92\%) & 77.14\% ($\uparrow$3.44\%) & 47.92\% ($\uparrow$6.89\%) & 63.46\% ($\uparrow$5.07\%) \\
    CodeGPT & 57.69\% ($\uparrow$11.09\%) & 77.70\% ($\uparrow$2.69\%) & 48.13\% ($\uparrow$6.42\%) & 64.08\% ($\uparrow$4.06\%) \\
    \midrule
    CodeT5 (\toolname{}) & \textbf{64.09\%} & \textbf{79.79\%} & \textbf{51.22\%} & \textbf{66.68\%} \\
    \bottomrule
    \end{tabular}
    \label{tab:rq3}
\end{table*}

\textbf{\emph{Results and Analysis.}}
Table~\ref{tab:rq3} presents the comparison performance of \toolname{} with different PLMs as assertion generators.
Overall, {\toolname} consistently achieves impressive performance across all PLMs and two datasets with an average accuracy of 54.29\% and CodeBLEU score of 70.88\%.
Particularly, when comparing the performance among different PLMs, we find the default generator of \toolname{} (\ie CodeT5) achieves better performance than the other four PLMs for all metrics and datasets.
For example, on the \olddata{} dataset, the prediction accuracy improvement against CodeBERT, GraphCodeBERT, UniXcoder, and CodeGPT reaches 6.37\%, 8.32\%, 8.92\%, and 11.09\%, respectively.
Similarly, CodeT5 outperforms other PLMs by an average of 6.95\%, 3.62\% and 4.90\% for other three settings, \ie accuracy on \olddata{}, accuracy, and CodeBLEU on \newdata{}.
The possible reason for the advance of CodeT5 lies in the model architecture.
Similar to \atla{}, CodeT5 features an encoder-decoder Transformer architecture, which has been shown to effectively support code generation tasks in prior work~\cite{wang2021codet5}.
However, encoder-only PLMs (such as GraphCodeBERT) necessitate a decoder for generation tasks, where the decoder starts from scratch and does not leverage the pre-training dataset.
Thus, it is natural and effective to employ CodeT5 as the backbone of \toolname{}.

When comparing different PLMs against previous AG approaches (shown in Table~\ref{tab:rq1}), we find that all PLMs are able to achieve impressive performance.
For example, five investigated PLMs achieve 57.69\%$\sim$64.09\% and 46.96\%$\sim$51.22\% prediction accuracy on \olddata{} and \newdata{}, outperforming the most recent baseline \edit{} by 7.91\%$\sim$19.88\% and 5.86\%$\sim$15.46\%, respectively.
Our analysis reveals that the observed improvements over previous baselines primarily stem from both our assertion generator and retriever, as demonstrated in Fig.~\ref{fig:case} and Fig.~\ref{fig:case2}.
The generator leverages extensive codebases to learn more meaningful vector representations for unit tests (\eg the pre-training data of CodeT5 comprises 2.3 million functions from \revise{CodeSearchNet~\cite{lu2021codexglue}}), whereas baselines are trained only on a restricted corpus of test-assert pairs.
Besides, the retriever identifies relevant assertions with both lexical and semantic matching, which is valuable for guiding the generator in generating correct assertions.

\finding{3}{Our comparison results demonstrate that:
(1) \toolname{} is general to different PLMs and sustainability achieves state-of-the-art performance, \eg five involved PLMs achieve an accuracy of 54.29\% on two datasets, outperforming the most recent \edit{} by 10.86\% on average;
(2) the default assertion generator (\ie CodeT5) is natural and quite effective in the unit assertion generation scenario, \eg improving the accuracy of other PLMs by 7.81\% on average across two datasets.
}

\section{Discussion}
\label{sec:discussion}
\begin{figure}[t]
	\graphicspath{{figs/}}
	\centering
	\includegraphics[width=0.7\linewidth]{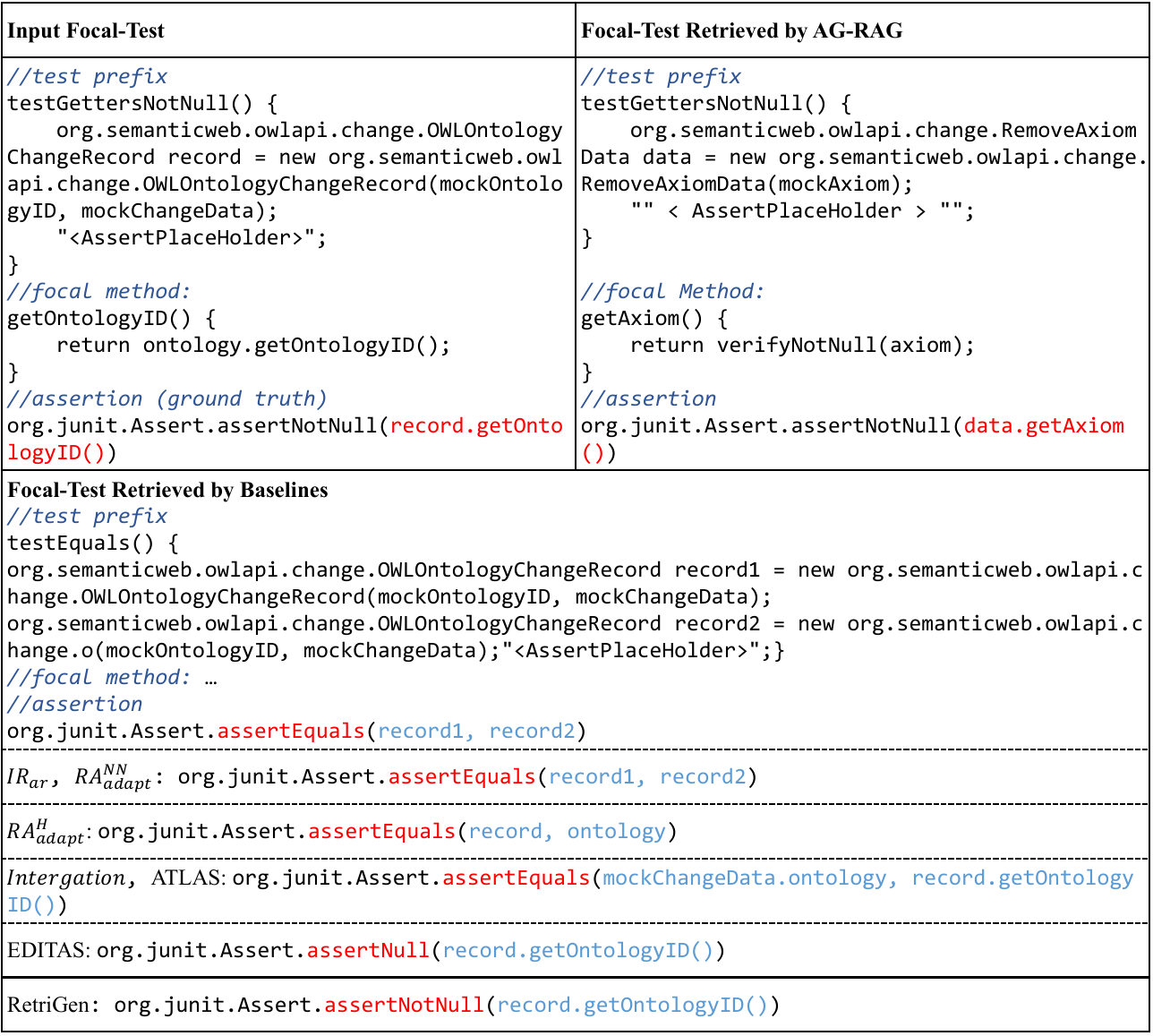}
	\caption{Example of assertions generated by AG approaches from OWLAPI}
    \Description{Example of assertions generated by AG approaches from OWLAPI}
	\label{fig:case}
\end{figure}

\textbf{Case Study}.
To further illustrate the retrieval and generation capabilities of \toolname{}, we present two examples of assertions from real-world projects.
Fig.~\ref{fig:case} illustrates a unique assertion from the OWLAPI project~\cite{owlapi}, which is only correctly generated by \toolname{}, but all baselines fail to.
In this example, \irar{} retrieves similar assertions based on lexical matching, and returns an assertion within the same class as the input focal-test, \ie ``OWLOntologyChangeRecord''.
Although the retrieved assertion has a high token similarity with that of the input focal-test (\eg both containing ``org.semanticweb.owlapi.change.OWLOntologyChangeRecord'' and ``record''), they are not responsible for testing similar functionalities.
Thus, \irar{} and \rann{} directly return the wrong assertion.
\rah{} and \edit{} also fail to produce correct assertions as they make modifications to the wrong assertion type.
For example, \rah{} attempts to replace the second parameter within the assertion (``record2'' $\rightarrow$ ``ontology'').
In contrast, {\toolname}, which relies on lexical and semantic matching, accurately identifies a similar assertion from another class file.
Despite significant differences in lexical similarity, the two assertions share the same assertion type (\ie assertNotNull) and parameter setting (\ie objectInstance.method()).
Thus, \toolname{} is able to capture the edit patterns between the two focal-tests, and performs the appropriate modifications on the retrieved assertion to generate the correct assertion.

\begin{figure}[htbp]
	\graphicspath{{figs/}}
	\centering
	\includegraphics[width=0.7\linewidth]{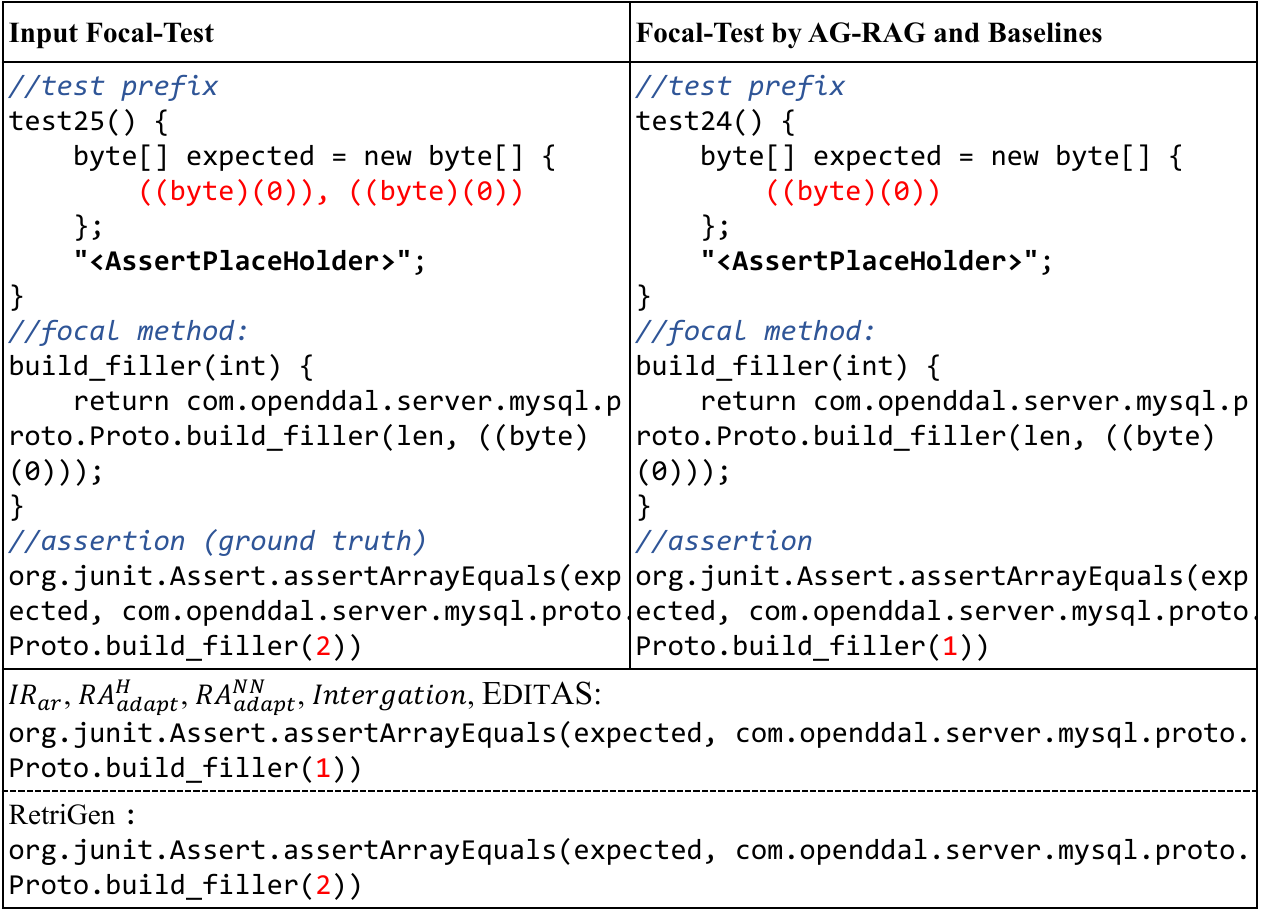}
	\caption{Example of assertions generated by AG approaches from OpenDDAL}
    \Description{Example of assertions generated by AG approaches from OpenDDAL}
	\label{fig:case2}
\end{figure}

Similarly, Fig.~\ref{fig:case2} presents a real-world example from OpenDDAL~\cite{openddal}, in which \toolname{} and all baselines retrieve the same assertion for the given focal-test.
The retrieved assertion is almost accurate, with only one parameter being corrected, as it targets the same focal-method (``build\_filler()'') with the query input.
However, existing approaches fail to assume this retrieved assertion is correct and directly return it as the final output.
Benefiting from the code understanding capability of the utilized PLM-based generator, \toolname{} successfully captures the semantic differences between the retrieved test prefix and the input test prefix (``\{((byte)(0))\}'' $\rightarrow$ ``\{((byte)(0)),((byte)(0))\}''), and applies the corresponding edit operations (``1'' $\rightarrow$ ``2'') to generate the correct assertion.

\textbf{Potential of Large Language Models}.
As mentioned in Section~\ref{sec:approach}, we consider encoder-decoder PLMs as foundation models of \toolname{} and select CodeT5 because it is quite effective and the most popular PLM that is fine-tuned to support sequence-to-sequence code generation tasks.
We notice that recent LLMs have been released with powerful performance, such as CodeLlama~\cite{roziere2023code}.
Most prior studies employ such LLMs in a zero-shot or few-shot setting, as fine-tuning these models with billions (or even more) of parameters is unaffordable due to device limitations~\cite{wang2024software}.
In this section. we attempt to explore the preliminary potential of integrating LLMs with \toolname{}.
Due to device limitations, we select CodeLlama-7B as the foundation model to generate assertions without using retrieved assertions.
The results demonstrate that CodeLlama-7B is able to achieve 71.42\% accuracy, which is 11.44\% better than \toolname{} (64.09\%).
It is worth noting that the improvement is valuable as we do not perform the retrieval-augmented process, while our hybrid retriever can significantly enhance the prediction results, \eg an improvement of 9.16\% on the \olddata{} dataset in Table~\ref{tab:rq2}.
Thus, we are confident that \toolname{} is able to achieve better results when equipped with recent LLMs.
\delete{Overall, the promising results motivate us to further explore the potential of \toolname{} with newly-released billion-level LLMs in the future.}
\revise{
While the potential of LLMs like CodeLlama is evident, their deployment raises concerns about computational efficiency in real-world applications.
For example, fine-tuning billion-level LLMs may require access to high-performance GPU clusters, making the process unaffordable for many research and industrial teams.
Thus, in the future, researchers can systematically analyze the trade-offs between performance gains and computational costs when integrating larger LLMs into \toolname{}.
Besides, it is crucial to employ optimization strategies (parameter-efficient fine-tuning) to reduce computational overhead without sacrificing significant performance.
Overall, the promising results motivate us to further explore the capabilities of \toolname{} with newly-released LLMs while balancing trade-offs between effectiveness gains and efficiency.
}

\textbf{Comparison with CEDAR}.
As mentioned in Section~\ref{sec:baselines}, following \revise{the methodology of} the most recent work \edit{}~\cite{sun2023revisiting}, we select six prior AG approaches from three categories as baselines.
To the best of our knowledge, this represents the largest set of baselines in the deep assertion generation literature.
Despite that, we notice that there may exist some studies leveraging LLMs in generating unit assertions~\cite{nashid2023retrieval}.
For example, Nashid~\etal~\cite{nashid2023retrieval} design a prompt-based few-shot learning strategy, CEDAR, which queries Codex for both assertion generation and program repair.
We exclude CEDAR as a baseline in Section~\ref{sec:results} mainly due to the significant data leakage issue of the utilized black-box LLM Codex.
\delete{Thus, in this section, we attempt to explore how \toolname{} performs in comparison to CEDAR.}
\revise{In this section, we conduct an extended experiment to explore how \toolname{} performs in comparison to CEDAR.}
However, we could not include CEDAR directly as a baseline because the datasets used in our study differ from those in theirs. 
CEDAR is evaluated on the \atla{} dataset, while our work and all baselines in Section~\ref{sec:baselines} are evaluated on \olddata{} and \newdata{}. 
In particular, \olddata{} and \newdata{} are constructed based on \atla{}, where \olddata{} removes certain samples (\eg those that could not be tokenized), and \newdata{} adds samples (\eg those with unknown tokens).
\delete{To address this, we conduct an additional experiment, comparing \toolname{} and CEDAR on the \atla{} dataset.}
\revise{To address this, we reuse the reported results of CEDAR and execute \toolname{} on the same \atla{} dataset for a direct comparison.}
Given that CEDAR uses Codex, a 12-billion-parameter model, and Codex is not open-source, we chose CodeLlama-7B as the foundation model for generating assertions in \toolname{}.
\delete{We find that \toolname{}, equipped with CodeLlama-7B, achieves 75.30\% accuracy on the \atla{} dataset, comparable to CEDAR's 75.79\% and 76.55\% in two settings.}
\revise{The results demonstrate that, equipped with CodeLlama-7B, \toolname{} is able to achieve 75.30\% accuracy on the \atla{} dataset, comparable to CEDAR's 75.79\% and 76.55\% in two settings.}
It is important to note that CEDAR relies on six retrieved assertions to guide Codex in generating assertions, while we fine-tune CodeLlama-7B with only one assertion. 
Compared to PLMs, LLMs have the capability to handle longer inputs, allowing the number of retrieved examples to significantly enhance prediction results, which is also proven in the original paper of CEDAR, \eg only 44.41\% accuracy of CEDAR with the one-shot setting.
Thus, we believe that \toolname{} could achieve even better results with more retrieved assertions.
In the future, we intend to undertake more comprehensive experiments with more retrieved assertions.

\revise{
In fact, \toolname{} and CEDAR are orthogonal, representing the cutting-edge approaches in model utilization strategies: fine-tuning and few-shot learning.
In this work, we implement \toolname{} in a fine-tuning manner instead of few-shot learning primarily due to concerns about data leakage.
Particularly, few-shot typically relies on a highly powerful foundation model with a record-breaking number of parameters, such as ChatGPT and Codex.
However, such commercial models are usually trained on publicly available data from the internet to incorporate extensive domain knowledge to address various domain-specific issues (\eg unit assertions), which may raises significant risks about data leakage.
In contrast, the fine-tuning strategy enables us to choose much smaller open-source models that can be more effectively adapted to specific domains while mitigating data leakage concerns.
For instance, in our work, \toolname{} in implemented with CodeT5, a model with only 220 million parameters, which achieves impressive performance compared to the latest EditAS approach. 
Despite its smaller size (just 1.8\% of Codex's parameters), CodeT5 effectively addresses data leakage concerns by relying entirely on open-source model weights and datasets.
Besides, \toolname{} is particularly valuable in real-world scenarios, such as where deployment resources are limited or where commercial models are unsuitable due to confidentiality concerns.
This makes \toolname{} a practical and reliable solution for tasks requiring domain-specific fine-tuning without compromising data security.
}

\textbf{Potential of Fault Detection Capabilities}.
As mentioned in Section~\ref{sec:metrics}, we employ two metrics to evaluate the performance of \toolname{} and baselines, which are widely adopted in prior deep assertion generation studies~\cite{he2024empirical}.
In this section, we attempt to investigate the potential of generated assertions in uncovering real-world bugs.
Following prior studies~\cite{dinella2022toga,liu2023towards,hossain2023neural}, we utilize EvoSuite~\cite{fraser2011evosuite} to generate test cases for Defects4J~\cite{just2014defects4j}, which includes 835 bugs from 17 real-world Java projects. 
Given EvoSuite’s reliance on randomized algorithms, we execute it 10 times per bug with different seeds to generate the final test cases.
We exclude test cases that involve exception behavior, focusing specifically on assertion generation. 
We then remove assertion statements generated by EvoSuite and apply \toolname{} to predict the test assertions.
To assess the performance of bug detection, we run the complete test cases on both buggy and fixed versions of all programs. 
A bug is considered detected if the test case fails on the buggy version but passes on the fixed one.
We compare \toolname{} with the best-performing baseline \edit{} due to dynamic execution overhead and computational resources. 
Both \toolname{} and \edit{} are implemented using \newdata{} as the training and retrieval corpus to ensure a fair comparison. 
Our results show that \toolname{} and \edit{} detect 33 and 21 real-world bugs, respectively, with 20 and 8 unique detections. 
\toolname{} surpasses \edit{} by 57.14\% in total detected bugs and by 150\% in unique detections. 
To our knowledge, this is the first attempt to evaluate the fault detection capabilities of deep assertion approaches~\cite{sun2023revisiting,nashid2023retrieval,yu2022automated,watson2020learning}, highlighting \toolname{}'s potential in detecting bugs. 
In the future, we recommend that researchers conduct more extensive studies with real-world bugs to evaluate deep assertion generation approaches.

\textbf{Practical Deployment}.
\revise{
As shown in Fig.~\ref{fig:workflow}, \toolname{} is designed to predict appropriate assertion statements when provided with focal methods and test prefixes.
In practice, \toolname{} can be deployed as a complement to automated test case generation tools or as a code completion plugin for developers.
First, existing automated test generation tools (\eg EvoSuite) are skilled at generating test prefixes with high code coverage but often struggle to understand the semantics of focal methods and generate meaningful assertions.
\toolname{} can address this gap by accepting a test prefix generated by existing tools along with its focal method as inputs and generating appropriate assert statements.
In this context, \toolname{} is not a replacement for existing automated test generation tools but a complementary technique that enhances their effectiveness.
Our preliminary results on Defects4J demonstrate that when provided with focal methods and the test prefixes generated by EvoSuite, AG-RAG can generate meaningful assertions capable of detecting real-world bugs.
Second, \toolname{} can be integrated into an IDE as a code completion plugin for developers, offering suggested assertion statements while they write test code. 
For instance, when a developer needs to validate the correctness of software units, they begin by writing a test prefix, \ie a sequence of call statements that invoke the specific behavior of the unit under test. 
\toolname{} then assists by automatically generating the necessary test assertions based on the context of the focal method and the test prefix. 
This plugin is particularly valuable because manually crafting effective assert statements requires a deep understanding of the program's functionality and specifications. 
Our experimental results on \olddata{} demonstrate that \toolname{} predicts 64.09\% of correct assertions when given the focal method under test and a developer-written test prefix. 
This result highlights the potential of \toolname{} in automated code completion scenarios, where developers can easily select appropriate assert statements recommended by \toolname{}, significantly reducing the manual effort involved in writing assertions.
}

\section{Threats to Validity}
\label{sec:threats}
\delete{The first threat to validity pertains to the evaluation metrics.
Following all of our baselines~\cite{watson2020learning,yu2022automated,sun2023revisiting}, we employ prediction accuracy to measure the effectiveness of the generated assertion.
We cannot compile these assertions due to the constraints of the utilized datasets that only contain code snippets of focal-tests.
Besides, automated building has consistently been a challenging task, heavily reliant on external/internal resources/dependencies~\cite{lou2020understanding}.
To mitigate the threat, we include a code-aware metric, CodeBLEU, that has not yet been adopted in prior AG work.
We also utilize Defects4J to calculate the number of real-world bugs detected by deep assertion generation approaches in a more realistic assessment setting. 
In the future, we will evaluate assertions in terms of compilability with more comprehensive datasets.}

\revise{The first threat to validity pertains to the evaluation metrics.
In this work, we strictly follow the evaluation protocols established in prior work~\cite{watson2020learning,yu2022automated,sun2023revisiting} and employ prediction accuracy to measure the effectiveness of the generated assertion.
However, we are unable to execute generated assertions due to the limitations of the utilized datasets (only containing focal and test methods)~\cite{lou2020understanding}, making it impossible to employ dynamic metrics such as defect detection rate.
To mitigate this threat, we utilize Defects4J to assess whether the generated assertions can detect real-world bugs.
Besides, since all baselines omit efficiency metrics, such as training or inference times in their evaluation, we cannot provide a direct comparison with baselines in this regard. 
To mitigate this threat, we include a code-aware metric, CodeBLEU, that has not yet been adopted in prior AG work.
In the future, we will conduct an extensive evaluation of such deep assertion generation approaches in terms of computational costs.}

The second threat to validity arises from the possibility of data leakage in PLMs.
PLMs are usually pre-trained with a mass of open-source projects, which may contain the test methods from the two AG datasets.
To address the concern, we query the pre-training datasets (\eg \revise{CodeSearchNet~\cite{lu2021codexglue}}) of studied PLMs and find such PLMs do not have access to any test cases (including assertions) during pre-training.
\revise{It is worth noting that the data leakage concern motivates our choice of open-source PLMs (mentioned in Section~\ref{sec:rq3}) instead of black-box PLMs or LLMs.}
Therefore, we believe that this concern does not significantly impact our conclusions.

\delete{
The third threat to validity is the potential limitation of our findings in generalizing to other \revise{benchmarks,} programming languages and PLMs.
First, we conduct experiments on two Java datasets.
We believe that the impact of this threat is relatively minor because 
(1) Java is the most targeted language in the AG field~\cite{he2024empirical}; 
(2) adopted datasets are widely-utilized and sufficiently large-scale to yield reliable conclusions; 
(3) our retriever is language-agnostic without any code-specific features; and (4) our PLM-based generator can naturally support multiple languages.
Second, we implement \toolname{} with the recent PLM CodeT5.
To mitigate the threat, we select four other PLMs from three types of model architectures.
It is important to note that larger LLMs have been recently released, such as Code~Llama~\cite{roziere2023code}.
However, most prior studies employ such LLMs in a zero-shot or few-shot setting, as fine-tuning these models with billions (or even more) of parameters is unaffordable due to device limitations~\cite{wang2024software}. 
Considering that selected PLMs cover different architectures, organizations, parameter sizes, and pre-training datasets, we are confident in extending our findings to newly released larger LLMs.
In the future, we attempt to explore these newly-released larger LLMs in assertion generation.}

\revise{
The third threat to validity is the potential limitation of our findings in generalizing to other benchmarks, programming languages, and PLMs.
First, the two benchmarks may not fully capture the diversity of assertion patterns or coding styles across various domains.
To mitigate the threat, we also utilize Defects4J to evaluate the fault detection capabilities of \toolname{} in a more realistic assessment setting. 
Second, we only evaluate \toolname{} on Java programs.
However, we believe that the impact of this threat is relatively minor because 
(1) Java is the most targeted language in the AG field~\cite{he2024empirical}; 
(2) adopted datasets are widely-utilized and sufficiently large-scale to yield reliable conclusions; 
and (3) \toolname{} is language-agnostic to support multiple languages naturally.
Third, we implement \toolname{} with the recent PLM CodeT5.
To mitigate the threat, we select four other PLMs from three types of model architectures.
Considering that selected PLMs cover different architectures, organizations, parameter sizes, and pre-training datasets, we are confident in extending our findings to newly released larger LLMs.
In the future, we attempt to explore the performance of \toolname{} with more benchmarks, programming languages and PLMs.
}

\section{Conclusion and Future Work}
\label{sec:conclusion}

In this work, we present {\toolname}, a novel retrieval-augmented assertion generation (AG) approach that leverages the advances of external codebases and pre-trained language models (PLMs).
{\toolname} first builds a hybrid retriever to search the most relevant assertion for a query focal-test from external codebases by both the lexical and semantic similarity.
{\toolname} then utilizes off-the-shelf CodeT5 as the assertion generator to predict assertions fine-tuned with both focal-tests and additional retrieved assertions.
The experimental results on two datasets show the superior performance of \toolname{} against all six baselines on two metrics, \eg achieving 57.66\% and 73.24\% in terms of accuracy and CodeBLE, outperforming all state-of-the-art AG techniques by 50.66\% and 14.14\% on average.
We also demonstrate that {\toolname} is able to generate a mass of unique assertions that all baselines fail to generate, \eg 1598 and 1818 ones on two datasets, 3.71X and 4.58X more than the most recent technique \edit{}.
In the future, we plan to explore the applicability of \toolname{} with larger PLMs, other programming languages, and datasets.

\section*{Acknowledgments}
The authors would like to thank the editors and anonymous reviewers for their time and comments. This work is partially supported by the National Natural Science Foundation of China (U24A20337, 61932012, 62372228), ShenzhenHong Kong-Macau Technology Research Programme (Type C) (Grant No. SGDX20230821091559018), CCF-Huawei Populus Grove Fund (CCF-HuaweiSE202304, CCF-HuaweiSY202306) and the Fundamental Research Funds for the Central Universities (14380029).

\bibliographystyle{ACM-Reference-Format}
\bibliography{reference}

\end{document}